\documentclass[openacc]{rstransa}

\usepackage{cite}
\usepackage{xcolor}
\usepackage{bm}

\titlehead{Research}
        
\begin{document}

\title{Constraining dark matter halo profiles with symbolic regression}


\author{
Alicia Martín,$^{1}$
Tariq Yasin,$^{1}$
Deaglan J. Bartlett,$^{2,1}$
Harry Desmond$^{3}$
and Pedro G. Ferreira$^{1}$}

\address{$^{1}$Astrophysics, University of Oxford, Oxford, OX1 3RH, United Kingdom\\
$^{2}$CNRS \& Sorbonne Université, Institut d’Astrophysique de Paris (IAP), UMR 7095, 98 bis bd Arago, F-75014 Paris, France\\
$^{3}$Institute of Cosmology \& Gravitation, University of Portsmouth, Portsmouth, PO1 3FX, United Kingdom}

\subject{astrophysics, machine learning}

\keywords{dark matter, symbolic regression}

\corres{Alicia Martín\\
\email{alicia.martin@physics.ox.ac.uk}}

\begin{abstract}

Dark matter haloes are typically characterised by radial density profiles with fixed forms motivated by simulations (e.g. NFW). However, simulation predictions depend on uncertain dark matter physics and baryonic modelling. Here, we present a method to constrain halo density profiles directly from observations using Exhaustive Symbolic Regression (ESR), a technique that searches the space of analytic expressions for the function that best balances accuracy and simplicity for a given dataset. We test the approach on mock weak lensing excess surface density (ESD) data of synthetic clusters with NFW profiles. Motivated by real data, we assign each ESD data point a constant fractional uncertainty and vary this uncertainty and the number of clusters to probe how data precision and sample size affect model selection. For fractional errors around 5\%, ESR recovers the NFW profile even from samples as small as $\sim 20$ clusters. At higher uncertainties representative of current surveys, simpler functions are favoured over NFW, though it remains competitive. This preference arises because weak lensing errors are smallest in the outskirts, causing the fits to be dominated by the outer profile. ESR therefore provides a robust, simulation-independent framework both for testing mass models and determining which features of a halo’s density profile are genuinely constrained by the data.

\end{abstract}

\maketitle

\section{Introduction}\label{sec:intro}

In the $\mathrm{\Lambda}$CDM cosmological model, structure forms hierarchically: small initial overdensities grow over time through gravitational instability, merging to form larger systems until they reach virial equilibrium. This process primarily affects dark matter, which interacts only through gravity. As a result, dark matter is the first component to form gravitationally bound structures known as dark matter haloes, creating a scaffolding into which baryons, subject also to dissipative processes, later settle. Understanding the properties and dynamics of haloes is important for studying the formation of galaxies and testing the predictions of cosmological models~\cite{frenk1988formation, blumenthal1984formation}.

Gravity-only $N$-body simulations trace the formation and evolution of dark matter haloes, while hydrodynamical simulations also attempt to include baryonic processes. From these simulations, a key result is that haloes seem to follow increasingly steep central density profiles or \emph{cusps}. This behaviour is commonly captured by the Navarro-Frenk-White (NFW) density profile ~\cite{navarro1997universal}:
\begin{eqnarray}
    \rho_{\text{NFW}}(r) = \frac{\rho_0}{r/r_s( 1 + r/r_s)^2} \quad,
    \label{eq:NFW}
\end{eqnarray}
where $r$ is the radial distance from the centre of the halo, $\rho_0$ is a characteristic density and $r_s$ a scale radius. 

Despite this profile being widely used, relying only on numerical simulations has limitations. These profiles are traditionally derived as empirical fits to simulations rather than predictions from first principles, and therefore lack a fundamental theory explaining their origin. Recently, progress has been made in deriving halo profiles based on the physics of collisionless relaxation, offering an insight on why this profiles appear in cold dark matter simulations \cite{banik2025collisionless}. However, the lack of a fully settled theory limits the reliability of these models. More significantly, observational  data continue to challenge the universality of cuspy profiles.
For instance, galaxy rotation curves often suggest flatter inner density slopes, so-called \emph{cores}, with an approximately constant density near the centre ($\rho \propto  r^0)$~\cite{de2010core}.

Several mechanisms have been proposed to explain the discrepancy between simulations and observations. Most involve baryonic processes, such as supernova-driven feedback expelling gas from the centre, or clumps transferring energy outwards through dynamical friction~\cite{duffy2010impact, schaye2015eagle}. However, these processes are very complex and remain difficult to model accurately in simulations. Moreover, some of these relevant processes occur below the resolution limit of such simulations, necessitating uncertain, semi-analytic ``subgrid models''. Beyond baryons, alternative dark matter models, such as fuzzy or self-interacting dark matter~\cite{vogelsberger2014properties}, can also generate halo profiles that differ from those seen in cold dark matter simulations. The space of theoretical predictions across both varying dark matter and baryonic physics is thus vast and difficult to constrain. 

Upcoming surveys will provide higher-precision measurements of dark matter distributions in galaxies and galaxy clusters. For example, data from surveys like Euclid \cite{Euclid_2011}, DESI \cite{DESI_2016}, and Rubin \cite{LSST_2009} will make it possible to combine weak lensing and dynamical data to reconstruct full 3D density profiles of galaxies and clusters. To make full use of these data, and to interpret them without imposing prior assumptions from simulations, we require a model-independent method that can uncover the structure of haloes directly from observations.

To do this, we propose using a machine learning algorithm called Symbolic Regression (SR) \cite{Kronberger_2024}. SR searches the space of analytic functions to find a mathematical expression that can describe a dataset well.  This makes it particularly suitable for situations like ours where we do not want to assume any specific functional form from the outset. In this way, it can infer dark matter halo density profiles from observations without relying on assumptions from simulations.

One of the main advantages of SR over other machine learning methods is that it produces interpretable mathematical expressions instead of black-box models. While it may be too optimistic to expect that these expressions will directly reveal new physical laws, they can still serve as models for testing correlations between profile parameters or for identifying systematic deviations from the standard halo profiles. Moreover, because these profiles are more accurate empirically, they can also be dropped into existing analyses (for example, when making predictions using the halo model) as a replacement for the NFW profile, providing an observationally better-motivated alternative.

In this work, we focus on a specific SR algorithm called Exhaustive Symbolic Regression (ESR) \cite{bartlett_exhaustive_2024}. Given a set of mathematical operators, ESR generates all possible combinations to produce functions up to a given level of complexity (defined to be equal to the number of nodes in the tree representation of a function). For each candidate function, it assesses how well it fits the data and how simple it is. This trade-off is formalised using the Minimum Description Length (MDL) principle: essentially, the best model is the one that gives a good fit to the data using the simplest possible formula.

The paper is structured as follows: we begin by introducing the ESR algorithm and describing how it can be used to combine data from multiple datasets (Sec. \ref{sec:methods}). Then we describe how it can be used to constrain halo profiles from weak lensing data. In Sec. \ref{sec:results} we apply ESR to mock weak lensing data to illustrate its effectiveness and discuss how data quality affects the resulting functions. Finally, we conclude and outline how the method can be extended to other probes such as galactic rotation curves in Sec. \ref{sec:conclusion}. Throughout this paper we adopt a concordance $\Lambda$CDM cosmology with $\Omega_{\mathrm{m}} = 0.27$, $\Omega_{\Lambda} = 0.73$ and $H_0 = 70\,\mathrm{km\,s^{-1}\,Mpc^{-1}}$ and use $\log$ to denote the natural logarithm.

\section{Methods}
\label{sec:methods}

\subsection{Symbolic regression framework}
\label{sec:ESR}

Symbolic Regression (SR) is a machine learning framework designed to recover mathematical expressions directly from data. Given a dataset and likelihood model, SR aims to find the equations that best describe the underlying relationships~\cite{bartlett_exhaustive_2024, desmond_functional_2023, tenachi2024class}. The goal is to balance accuracy, ensuring the final expression reproduces the data reliably, and simplicity, to find equations that are easy to interpret and more likely to generalise well. In doing so, SR can produce insight into the physical processes that generated the data.

In recent years, many SR algorithms have been developed. Most begin by specifying a set of mathematical operators (e.g. $+$, $-$, $\times$, $\exp$, etc.) and proceed by combining these to generate candidate expressions. The complexity of an expression is defined as the number of operators, parameters and variables needed to construct the function. The most common approaches to generating functions at different complexities rely on genetic programming, which evolves the initial population of functions through operations such as mutation and crossover\cite{turing2009computing, goldberg1994genetic, cranmer_interpretable_2023}. These methods aim to be computationally efficient by only studying a subset of the function space. The hope is that this is the optimal subset for the data in question, however this exploration often misses important expressions even at relatively low complexities.~\cite{bartlett_exhaustive_2024, kronberger2024inefficiency}

This limitation motivated the development of Exhaustive Symbolic Regression (ESR), an algorithm that, given a set of operators, generates all possible functions up to a specified complexity~\cite{bartlett_exhaustive_2024, sousa_optimal_2024, desmond_functional_2023}. This brute-force approach constructs and tests every valid expression within the allowed function space, ensuring no potentially good functions are missed. This makes ESR the best algorithm to find expressions at low complexities and a valuable benchmark for guiding stochastic methods such as genetic programming at higher complexity.

Although ESR is described in detail in \cite{bartlett_exhaustive_2024}, in this section we briefly describe its two main stages. First, it systematically generates all possible expressions at a given complexity and optimises their free parameters. Second, it ranks the resulting expressions using an information theory-motivated metric which accounts for both goodness-of-fit and model complexity.

\subsubsection{Generating and optimising functions}

To generate mathematical expressions, ESR begins by the user specifying a set of operators. These can be classified into three types: binary (e.g. $+$, $\times$), unary (e.g. $\exp$, $\log$) and nullary (parameters and variables). Mathematical expressions can be represented as trees, where each node corresponds to an operator and the structure of the tree encodes the mathematical composition of the function~\cite{petersen2019deep}. The complexity of a function can also be defined as the number of connected nodes in the tree representation. For example, the standard NFW profile can be written as $\theta_0 / (r (|\theta_1| + r)^2)$, which is a complexity $9$ function (the modulus sign is not counted as an operator).

Given a fixed complexity, ESR systematically generates all possible tree structures with that number of nodes and then populates each node with every possible combination of operators with the correct arity. In this framework, the choice of operators and the maximum complexity are the only degrees of freedom.

Once the complete set of candidate expressions has been generated, ESR simplifies expressions and eliminates duplicates. Since multiple tree structures can generate equivalent mathematical expressions (e.g. $e^{a + x} = e^a\cdot e^x $ it is important to identify these cases as retaining only unique expressions avoids redundant computations in the next step.

Finally, ESR optimises the free parameters in each of the candidate models. We consider the best-fit parameters to be those that maximise the likelihood function. Here we use Gaussian likelihood with uncorrelated errors. We consider two optimisation algorithms to be particular useful for parameter optimisation. The faster option is to use a gradient-based method such as BFGS \cite{Broyden_1970,Fletcher_1970,Goldfarb_1070,Shanno_1970}, which uses gradient information to effectively navigate the likelihood surface and locate maxima. However, we find that this algorithm sometimes gets trapped in local maxima, particularly when the likelihood landscape is complex. To mitigate this, we also use a derivative-free method like \textsc{Nelder-Mead} \cite{Nelder_1965}, which can be more robust in avoiding local maxima, although it typically takes longer to converge. 

To ensure that the parameter estimation has converged to a reliable solution, the optimisation procedure is repeated up to $N_{\text{iter}}$ times. After each iteration we check whether the same best solution (within some likelihood tolerance) has been recovered $N_{\text{conv}}$ times, in which case we terminate the optimisation as successful. If the highest maximum has not been reached $N_{\text{conv}}$ times after $N_{\text{iter}}$ iterations, the algorithm returns the maximum-likelihood solution achieved. The values of $N_{\text{conv}}$ and $N_{\text{iter}}$ used here are the same as those specified in \cite{bartlett_exhaustive_2024}.

This process is repeated for all complexities considered. However, there is a practical upper limit to how high one can go. As the number of nodes is increased, the number of possible expressions grow exponentially, making the computation infeasible beyond a certain complexity. Since the standard NFW profile is a complexity $9$ function, we set our maximum search complexity to $9$, ensuring NFW is included in our analysis while remaining within these computational constraints.


\subsubsection{Model selection via Minimum Description Length}

The result of the previous procedure is a list of all functions up to a given complexity, along with their best-fit parameters. The next step is to rank these functions. A natural choice would be to use the maximum likelihood as a criterion. Plotting each function on the complexity-likelihood plane, the functions that achieve the highest likelihood at each complexity form what is known as the \textit{Pareto front}. Along this front, no function can be made more accurate without also becoming more complex and vice versa. 

However, the Pareto front itself only defines a set of optimal trade-offs; it does not provide a principled way of choosing one function over another within that set. In other words, Pareto optimisation gives a frontier of good functions in terms of accuracy but not a ranking. To obtain a one-dimensional ordering of the models, an additional criterion is needed that weights complexity against accuracy. To  do this, ESR uses an information-theory approach known as the \textit{Minimum Description Length} (MDL) principle~\cite{rissanen1978modeling, grunwald2007minimum, grunwald2019minimum}. MDL states that the best mathematical description of the data is the one that compress information the most, or in other words that needs the fewest units of information to transmit the data. 
Here we will perform all calculations with natural logarithms rather than base-2, and thus our unit of information is nats rather than bits.

This idea is represented by the \textit{Description Length}, $L(D)$, which is defined as $ L(D) = L(D|H) + L(H)$. Here, $L(D|H)$ quantifies how well the hypothesis $H$ fits the data $D$. Under the Shannon-Fano coding scheme, the optimal encoding for the residual has $L(D|H) = - \log \mathcal{L}(D|\hat\theta)$~\cite{cover1999elements}, where $\mathcal{L}$ is the likelihood and $\hat{\theta}$ are the maximum-likelihood parameters. The second term, $L(H)$, measures the complexity of the hypothesis itself. It penalises models with more operators, more parameters and parameters that need to be specified to higher precision to achieve a high likelihood.

The full expression for $L(D)$ is derived in \cite{bartlett_exhaustive_2024} and is given by
\begin{eqnarray}
    L(D) &=& - \log (\mathcal{\hat{L}}) + k\log(n) + p \log(2) + \sum_{i=1}^{p} \log(|\theta_i| / \Delta_i) + \sum_j \log(c_j),
\label{eq:DL}
\end{eqnarray}
where $k$ is the number of nodes to represent the function, $n$ is the number of unique operators, $p$ is the number of free parameters, $1 / \Delta_i$ is the precision to which the parameters can be specified
 and $c_j$ are constant natural numbers. The hat notation indicates evaluation at the maximum likelihood point.  

In some cases, it might happen that $|\theta_i|/ \Delta_i<1$, meaning that the ML estimate of a parameter is indistinguishable from $0$ within the uncertainty tolerance. In these situations, we set these parameters to $0$, recalculate the likelihood, and reduce the parameter count $p$ by one. Since this occurs only when a parameter is already poorly constrained, the overall effect on $L(D)$ is minimal.

Eq. \ref{eq:DL} can be interpreted as an approximation to the Bayesian evidence under a certain choice of parameter prior and function prior, where the latter is given by $L(H)$. More sophisticated choices for the prior can be expressed in terms of language models \cite{bartlett2023priors}, however, for simplicity, in this work we use the formulation of the MDL as given by \cite{bartlett_exhaustive_2024}.

This expression balances model simplicity and goodness of fit. Simple functions are penalised if they fail to fit the data well (high $L(D|H)$), while highly complex functions that overfit are penalised for their complexity (high $L(H)$). This is naturally influenced by the quality of the data. When the data are not particularly constraining (few data points or large uncertainties), the likelihood term has less influence, and simpler functions with higher prior probabilities (and hence low $L(H)$) are favoured.

\subsubsection{Global vs Local parameters}
\label{sec:global_params}

We aim to find dark matter density profiles that perform well across different galaxies or galaxy clusters. This requires combining results from individual systems to evaluate overall model performance. There are different ways to do this depending on how the model parameters are treated.

One option is to consider that the parameters are independent for each dataset, such that they are optimised separately for each galaxy or cluster. This means applying ESR independently on each system and afterwards combining the resulting fits. To compute the overall description length in this case, we sum the likelihood and parameter precision terms across galaxies, but count complexity terms (i.e. those depending on the number of nodes $k$ and operators $n$) only once. This results in a total description length given by

\begin{equation} 
    \begin{split}
        L(D) &= - \sum_{a=1}^N  \log \mathcal{L}^{(a)} + k\log(n) + N \cdot p \log(2) \\ 
        & + \sum_{a=1}^N \sum_{i=1}^p \log(|\theta_i^{(a)}| / \Delta_i^{(a)}) + \sum_j \log(c_j),
    \end{split}
    \label{eq:DL_global} 
\end{equation}
for $N$ independent galaxies/galaxy clusters labelled by the index $a$.

Alternatively, we can consider the possibility that a subset of parameters is shared across all galaxies; we refer to these as \textit{global} parameters. This allows for the possibility that some of the parameters are universal constants that appear in the mathematical description of dark matter haloes, for example the square in the denominator of the NFW profile. In this case, a single set of shared parameters is optimised using combined data from all systems. This results in a lower total likelihood, since the model has fewer degrees of freedom to fit each galaxy. However, it also reduces the model complexity by significantly decreasing the number of free parameters. This trade-off can lead the data to prefer global over local parameters.

Ideally, one would like to combine both approaches, allowing some parameters to be global while others remain galaxy-specific in the attempt to find the overall lowest description length models. This allows the data itself to determine which model structure is appropriate. To implement this in ESR, we use a two-step procedure. First, we perform fully local fits, where all parameters are independently fitted to each galaxy. This gives the best possible likelihood that each function can have.
Then, we evaluate whether $L(D)$ can be improved by ``globalising'' some of the parameters. The aim is to determine whether making some parameters global could potentially improve the $L(D)$ enough to improve the function's ranking beyond that of the best-performing function with all local parameters. To estimate this, we consider the maximum theoretical possible gain in $L(D)$, which occurs when the likelihood remains unchanged on globalisation. This yields (setting the precision cost of new global parameters to zero)

\begin{equation}
    \begin{split}
        \text{max} (\Delta L) &= \text{max} (L_{\rm \text{local}} - L_{\rm \text{global}}) \\
        &=
        \sum_{a=1}^N \sum_{i=1}^{p_{\rm \text{global}}} \log \left(|\theta_i^{(a)}| / \Delta_i^{(a)}\right) +  p_{\rm \text{global}} (N - 1) \log 2,
    \end{split}
\label{eq:globalise}
\end{equation}
where $p_{\text{global}}$ denotes the number of parameters being considered for globalisation. It is necessary to consider all possible combinations of local and global parameters for each function. 

Using this, we identify good candidates for globalisation as those for which $\text{max} (\Delta L)$ is large enough to put the function in the top $10$ ranked models. For any such function, we proceed to globalise the corresponding parameter and re-evaluate the description length. This involves re-fitting the function setting the desired parameters to global.


One straightforward way to do this is to perform a single optimisation over all $N_{\text{params}} = p_{\text{global}} + N \cdot p_{\text{local}}$ parameters, where $N$ is the number of galaxies or clusters. However, this requires searching in a very high-dimensional space, and we find that the optimiser often struggles to explore the likelihood surface effectively. 

We therefore use instead a nested optimisation. In this scheme, an outer optimiser searches in the space of global parameters. For each set of global parameters it tries, an inner optimiser fits the best corresponding local parameters for each cluster. This process is repeated until the global optimiser converges. We define convergence as $N_{\text{conv}}$ iterations showing an absolute improvement in the likelihood function of less than $0.1$.

While computationally more expensive, this method has the advantage that each optimisation, both for the local and global parameters, occurs in a low-dimensional space. This improves the convergence and stability of the fitting procedure. The rest of the terms for the description length are calculated as in Eq.~\ref{eq:DL_global}, where the sum is performed over local parameters and global parameters are only considered once.

\subsection{Application to weak lensing}
\label{sec:weak_lensing}

\subsubsection{Weak lensing basics}
Weak gravitational lensing refers to the magnification and distortion of background galaxy shapes caused by the deflection of light by intervening mass along the line of sight. It offers a powerful and direct way of mapping the matter distribution in galaxies and clusters without relying on assumptions about their dynamical state. Galaxy clusters, being among the most massive gravitationally bound systems, produce strong gravitational wells that make them particularity effective lenses. As such, they provide unique environments for studying the distribution of matter, which is dominated by dark matter.

The effects of weak gravitational lensing are described by the convergence $\kappa$, which describes the isotropic magnification, and the complex shear $\gamma$ that describes the anisotropic distortion of galaxy shapes. In practice, the observable quantity is not the shear $\gamma$ but the reduced shear $g$, which accounts for the magnification effect from the convergence

\begin{eqnarray}
g = \frac{\gamma}{1 - \kappa}.
\end{eqnarray}

The shear $\gamma$ can be decomposed into two components: the tangential component $\gamma_+$ and the $45^\circ$-rotated component $\gamma_{\times}$. At the radii we consider, \(g_+ \simeq \gamma_+\). The tangential shear $\gamma_+$ is directly related to the azimuthally averaged surface mass density $\Sigma(R)$ through~\cite{Bartelmann_2001}
\begin{equation}
    \label{eq:ESD_definition}
   \gamma_{+} = \frac{\langle\Sigma (<R)\rangle - \Sigma(R)}{\Sigma_{\text{cr}}(z_l, z_s)} \equiv \frac{\Delta \Sigma(R)}{\Sigma_{\text{cr}}(z_l, z_s)}, 
\end{equation}
where $R$ is the projected radius from the lens centre,  $\Delta \Sigma(R)$ is the Excess Surface Density (ESD) and $\langle\Sigma (<R)\rangle$ is the average surface density within $R$, given by
\begin{equation}
    \langle\Sigma (< R)\rangle = \frac{2}{R^2}\int_{0}^{R}{\Sigma(R^\prime ) R^\prime {\rm d} R^\prime }.
\end{equation}
The critical surface density $\Sigma_{\text{cr}}$ is
\begin{equation}
     \Sigma_{\text{cr}}(z_l, z_s) = \frac{c^2}{4 \pi G}\frac{D_{\rm A}(z_s)}{(1 + z_{\rm l})^2 D_{\rm A}(z_{\rm l}) D_A(z_{\rm l},z_{\rm s})},
\end{equation}
where $c$ is the speed of light, $G$ is the gravitational constant, $z_{\rm l}$ is the lens redshift, $z_{\rm s}$ is the source redshift, and $D_{\rm A}$ is the angular diameter distance. The factor $(1+z_{\rm l})^2$ converts to comoving surface mass density. 

The projected surface density, $\Sigma(R)$, can also be related to the three-dimensional mass density $\rho(r)$ by 
\begin{eqnarray}
    \Sigma (R) = \int_{-\infty}^{\infty}{\rho(\sqrt{R^2 + z^2}) dz},
\end{eqnarray}
\noindent where $z$ is the line-of-sight coordinate and $r$ is the three-dimensional distance, such that $r^2 = R^2 + z^2$. 


By combining these equations, we can derive the predicted $\Delta \Sigma (R)$ profile for any candidate three-dimensional density function $\rho (r)$. ESR can use this transformation to fit ESD data and rank the performance of each function. 
To simplify the numerical calculation of $\Delta \Sigma(R)$, we use an alternative expression that avoids double integrals~\cite{cromer_towards_2022}
\begin{equation}
     \Delta \Sigma (R) = \frac{4}{R^2} \int_0^R dr r^2 \rho(r) - 4R\int_0^{\pi/2} d\theta \frac{\rho(R \sec \theta)}{4 \sin \theta + 3 - \cos(2\theta)} .
    \label{eq:rho_to_esd}
\end{equation}




In principle, the total excess surface density $\Delta \Sigma (R)$ has contributions from several components: the one-halo dark matter term, baryonic components, and the two-halo term.

The baryonic contribution itself has two main parts: the stellar mass of the BCG, which is important only at very small scales ($R \lesssim 0.1 h^{-1}$~Mpc), and the hot, extended Intracluster Medium (ICM) or gas component. The two-halo term arises from correlated matter in surrounding large-scale structure and becomes important at scales approaching and beyond the virial radius \cite{umetsu_weak_2020}.

We analyse data in the intermediate radial range ($0.3 - 3 \, h^{-1} \text{Mpc}$), where the one-halo term dominates in $\Lambda\mathrm{CDM}$ for the cluster masses under consideration. We therefore neglect the two-halo term in our analysis, providing a detailed quantitative justification in Section \ref{sec:mock_data}. We do not model the baryonic components separately from the dark matter, with the hot ICM gas expected to contribute around 1/6th of the total mass. Therefore, by neglecting the gas component, this inference procedure applied to real data effectively constrains the total mass profile rather than the purely dark-matter contribution. We discuss how the method could be developed to explicitly model both 2-halo term and the subdominant baryon component in Section \ref{sec:conclusion}.

An important consideration is the quality of the data itself. Since more precise data tends to prefer more complex models, the quality of the data is important to understand if profiles like NFW can actually be recovered from observations. In the next section, we present a worked example using mock weak lensing data. We focus on the one-halo term only and investigate how data quality affects the performance of ESR, and the extent to which the function generating the data can be expected to be recovered by ESR.





\subsubsection{Mock data generation}
\label{sec:mock_data}

To illustrate the method and test the ability of ESR to recover known physical profiles, we generate a synthetic sample of $150$ galaxy clusters with an NFW profile, that is designed to closely resemble the dark matter halo properties inferred for the XXL X-ray selected cluster sample in the analysis of \cite{umetsu_weak_2020}. The halo masses were drawn from a Gaussian probability density function in $Z = \log (M_{200}/h^{-1} M_{\odot})$ with a mean $\mu_Z = 14$ and a dispersion $\sigma_Z = 0.5 \log 10$. The range of true $M_{200}$ masses for the simulated sample is $4.0 \times 10^{13} \lesssim M_{200}/(h^{-1} M_{\odot} ) \lesssim 5.0 \times  10^{14}$.  Corresponding concentrations were drawn from the scattered c$_{200}$-M$_{200}$ relation of \cite{bhattacharya2013dark}, assuming a log-normal intrinsic dispersion of $\sigma(\log c_{200}) = 0.15 \log 10$ and a fixed redshift of $z = 0.3$, the median redshift of the XXL sample.  

The true ESD, $\Delta \Sigma (R)$, profile for each cluster was calculated in $10$ logarithmically spaced bins of comoving radius from $R_{\text{min}}= 0.3 h^{-1}$ Mpc to $R_{\text{max}}= 3 h^{-1}$, again following \cite{umetsu_weak_2020}, avoiding the inner regions ($R < 0.3 h^{-1}$ Mpc). The lower limit is to ensure our measurements are not significantly affected by masking, imperfect deblending, or photometric biases caused by the bright cluster galaxies (BCGs). Furthermore, this $R_{\text{min}}$ is much larger than the typical offsets between the BCG and the X-ray peak for these clusters, meaning smoothing of the lensing signal due to miscentering is also not expected to be a significant effect. The upper limit is to avoid the region where the 2-halo term becomes significant.




To model observational uncertainties, we add realistic noise to the simulated measurements. For current surveys, the uncertainty in measured ESD profiles is dominated by the intrinsic dispersion of unlensed galaxy shapes. This \emph{shape noise} scales with the number of source galaxies in a radial bin ($n_i$) as $\sigma_i \propto 1/\sqrt{n_i}$. Because we use logarithmic radial bins, the annular area—and hence $n_i$—grows approximately as $R^2$, implying $\sigma_i \propto 1/R$. In real cluster data the ESD declines as approximately 1/R in the radial range we consider, yielding an approximately constant fractional uncertainty. Therefore we adopt the following error model for our synthetic data:
\begin{align}
\Delta\Sigma(R_i) &= \Delta\Sigma_{\rm true}(R_i) + \epsilon_i, \label{eq:meas_model}\\
\epsilon_i &\sim \mathcal{N}\!\left(0,\sigma_i\right), 
\qquad
\sigma_i = f\,\Delta\Sigma_{\rm true}(R_i), \label{eq:frac_sigma}
\end{align}
with $i=1,\ldots,10$ radial bins and $\mathcal{N}$ a normal distribution with mean zero and dispersion $\sigma_i$ . To test our method's performance under different data quality scenarios, we generated six mock datasets with
\[
f \in \{0.01,\,0.05,\,0.20,\,0.40,\,0.60,\,0.80\}.
\]

We can compare these $f$ values to typical fractional uncertainties of current lensing surveys. For instance, the HSC-XXL clsuter sample in \cite{umetsu_weak_2020} has a median fractional uncertainty of approximately $f \approx 0.60$, while a higher signal-to-noise survey like CLASH has an uncertainty closer to $f \approx 0.38$. \cite{umetsu2014clash}

Our chosen range, $f \in \{0.01, \ldots, 0.80\}$, therefore spans the full spectrum of data quality. The low-noise values ($f=0.01, f=0.05$) represent highly idealised precision, while the upper end of our range ($f=0.40, 0.60, 0.80$) is representative of, or even more challenging than, the data from current-generation surveys. This allows us to test our method's performance in both idealised and realistic observational scenarios.

Our analysis is restricted to modelling the one-halo term of the cluster mass profile. One might wonder whether the two-halo term contribution is significant, particularly at the outermost datapoints. In the $\Lambda$CDM framework, this question was addressed in detail by \cite{umetsu_cluster-galaxy_2020} for the HSC cluster sample. Since our dataset is designed to be comparable to the XXL-HSC sample, their findings are directly relevant. For their cluster sample, which has a lensing-weighted mean mass (the effective average mass of a cluster sample, obtained by weighting each cluster by the strength or precision of its lensing signal) of $M_{\rm WL} = 8.3 \times 10^{13} \, h^{-1} M_{\odot}$, they found the two-halo term's contribution to the stacked ESD profile to be negligible across their entire radial range (see their Fig. 4). Even at their outermost datapoint ($R \approx 3 \, h^{-1} \text{Mpc}$), the two-halo term was still an order of magnitude smaller than the fitted one-halo term.

We have confirmed this for our own sample by calculating the two-halo term (following the method in \cite{umetsu_cluster-galaxy_2020}) for our lowest-mass clusters, where this term is expected to be relatively most significant. For the lowest-mass cluster in our sample—where the 2-halo contribution is most significant—it remains about 4.5 times smaller than the 1-halo term at the outermost radial bin, and roughly 10 times smaller for the second-lowest-mass system

Given that the 2-halo term is so small, one route would be to simply include the standard $\Lambda$CDM 2-halo term as an additional source of uncertainty in the covariance. However, the above implies this would not noticeably alter our conclusions on the quality of data required to recover NFW. For a future application to real data, there is a deeper methodological issue: given the aim of the study is to recover the halo profile, assuming NFW to calculate the two-halo term is unsatisfactory. In principle, one could use each halo profile under consideration to calculate its own two-halo term, but this would be costly. In the near-term, the most practical option when applying this method to real data is to test the stability of the preferred profiles against the radial range of the data (e.g., by re-fitting with the outermost bins removed).


\section{Results}
\label{sec:results}

We applied our ESR pipeline to the mock cluster datasets, optimising all candidate functions up to complexity $9$ for each fractional uncertainty level. To illustrate the fitting procedure and its results, we will first focus on the low-noise regime ($f = 0.05$) as a representative example.

\begin{figure}
\includegraphics[width=\textwidth]{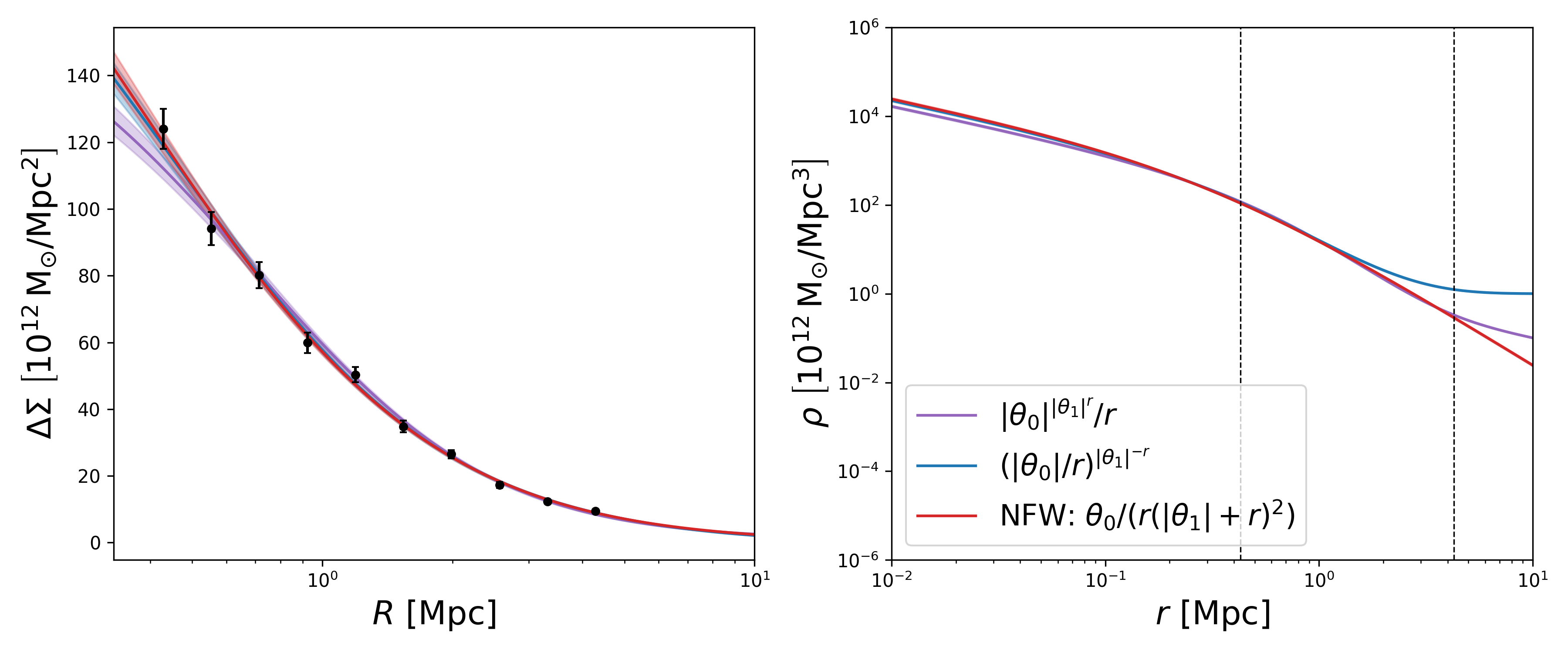}
\caption{An example result of the ESR fitting procedure for one mock galaxy cluster, generated with a fractional uncertainty of $f = 0.05$. \emph{Left panel:} The mock weak lensing ESD data (black points) are compared to the best-fit NFW profile (red) and the next two best-fitting functions 
discovered by ESR, $|\theta_0|^{|\theta_1|^r}/r$ (purple) and 
$\left(|\theta_0|/r\right)^{|\theta_1|^{-r}}$ (blue).
\emph{Right panel:} The corresponding 3D density profiles ($\rho$) for the three models as a function of the intrinsic 3D radius. The vertical dotted lines mark the radial range of 
    the ESD data from the left panel.} 
\label{fig:cluster}
\end{figure}

Fig. \ref{fig:cluster} shows the ESR-derived fits for a single mock cluster from this $f=0.05$ dataset. The left panel compares the mock ESD data (black points) to the best-fit NFW profile and the next two top-ranked functions discovered by ESR for this cluster, which are $|\theta_0|^{|\theta_1|^r}/r$ and $\left(|\theta_0|/r\right)^{|\theta_1|^{-r}}$. The right panel shows the corresponding 3D density profiles ($\rho$) for these same models.

We then ranked the functions based on their combined performance across all clusters. As explained in Section \ref{sec:global_params}, our procedure first optimises all functions assuming purely local parameters, and then identifies which models benefit from "globalising" one or more parameters. For this $f=0.05$ uncertainty level, this method identified $734$ functions as candidates for the globalising step out of the $36,217$ total unique functions evaluated up to complexity $9$ ($2\%$ of the functions). This results in a total of $2,432$ unique function and parameter-type (local/global) combinations that were then optimised.

Table \ref{tab:best_funcs} summarises the top-ranked expressions across all global and local combinations. The total description length for each function is broken down into three components: the residuals between the predicted and observed data, the structural complexity of the functional form, and the penalty due to the number of free parameters, as detailed in the table footnote.

\begin{table*}
    \centering
    \scriptsize
    \begin{tabular}{c c c c c c c c}
        \hline
        Rank & $\rho(r) / 10^{12} M_{\odot}\,\text{Mpc}^{-3}$ & Complexity & Global params & \multicolumn{4}{c}{Description length} \\
        \cline{5-8}
        &  &  &  & Residual$^{1}$ & Parameter$^{2}$ & Function$^{3}$ & Total \\
        \hline
        1 & NFW: $\theta_0 / (r (|\theta_1| + r)^2) $
        & 9 & -- & 618.40 & 1043.06 & 24.04 & 1685.51 \\
        2 & gNFW: $\theta_0 / (r (|\theta_1| + r)^{\theta_2}) $
        & 9 & $\theta_2$ & 617.79 & 1053.98 & 25.08 & 1696.85 \\
        3  & $\lvert \theta_{0}/r - r \rvert^{\,\lvert \theta_{1}\rvert^{\,r}}$  & 9 & -- & 647.02 & 1115.00 & 14.48 & 1776.50 \\
        4  & $(\lvert\theta_{0}\rvert/r)^{\,\lvert \theta_{1}\rvert^{-r}}$   & 8 & -- & 650.62 & 1115.01 & 12.88 & 1778.51 \\
        5  & $\theta_{0}/\big(r\,(\theta_{1}-r^{2})\big)$  & 9 & -- & 889.52 & 945.90  & 14.48 & 1849.90 \\
        6 & $\lvert \theta_0 \rvert^{\,\lvert 1/(\theta_1 - 1/r) \rvert^{\,r}}$ & 9 & -- & 902.78 & 1002.83 & 14.48 & 1920.09 \\
        7 & $\lvert \theta_0 \rvert^{\,\lvert \theta_1 \rvert^{-r}} / r$ & 8 & -- & 834.67 & 1076.40 & 12.88 & 1923.94 \\
        8 & $\lvert \theta_0 - r \rvert^{\,\lvert \theta_1 \rvert^{\,r}} / r$ & 9 & -- & 834.69 & 1076.15 & 14.48 & 1925.32 \\
        9 & $\lvert \theta_0 + r \rvert^{\,\lvert \theta_1 \rvert^{\,r}} / r$ & 9 & -- & 834.69 & 1076.18 & 14.48 & 1925.35 \\
        10 & $\theta_0 / (\theta_1 - r^{1-\theta_2})$ & 9 & -- & 570.25 & 1383.21 & 14.48 & 1967.94 \\
        \vdots & \vdots & \vdots & \vdots & \vdots & \vdots & \vdots & \vdots \\
        40 & gNFW: $\theta_0 / (r (|\theta_1| + r)^{\theta_2}) $
        & 9 & -- & 545.85 & 1597.07 & 25.08 & 2167.15 \\
        \hline 
    \end{tabular}
    \begin{tabular}{c}
        $^{1}\ -\sum_{a=1}^{N}\log\hat{\mathcal{L}}^{(a)}$ \qquad
        $^{2}\ p \log(2) + \sum_{a=1}^N \sum_{i=1}^p \log(|\theta_i^{(a)}| / \Delta_i^{(a)})$ \qquad
        $^{3}\ k\log n + \sum_{j}\log c_{j}$
    \end{tabular}
    \caption{Highest ranked density models ordered by their total description length
    $L(D) = \text{Residual} + \text{Parameter length} + \text{Function length}$. The ``Global params'' column specifies which parameters are set to global for each function.}
    \label{tab:best_funcs}
\end{table*}

The best-ranked function is the NFW profile. This is a validation of our mock-data simulation: in the low-noise regime, ESR successfully recovers the true, underlying model that was used to generate the data. The generalised NFW (gNFW) profile, which introduces an extra parameter $\theta_2$ per halo by allowing the inner slope of the profile to vary, achieves a better likelihood but is penalised for its extra complexity, falling to rank $40$ in total $L(D)$. We find a median and standard deviation for the slope $\theta_2$ of $2.01 \pm 0.31$ across all clusters, close to the expected NFW value of $2$.

The $2$nd-ranked function is the gNFW profile with its outer slope $\theta_2$ treated as a global parameter shared across all clusters. Its best-fit value of $\theta_2 = 2.024 \pm 0.008$ is extremely close to the canonical NFW value of $2$. This model achieves a slightly better likelihood (Residual $= 617.79$) than the standard NFW ($618.40$) but is penalised for having one extra parameter just enough by the MDL to fall to $2$nd place.

We also note that despite the global search, the final ranking is dominated by models with purely local parameters. The gNFW is the only function in this top list that successfully incorporates a global parameter. This makes sense, local parameters are necessary to capture cluster-to-cluster variations in mass and concentration, while global parameters are only appropriate for universal properties like an outer slope.

Interestingly, most of the highest-ranked functions have two free parameters, same as the standard NFW profile. The MDL metric heavily penalises additional parameters, and our results show that models with three or more parameters do not provide a sufficient improvement in fit to overcome this penalty.

\begin{figure}
\centering
\includegraphics[width=100mm]{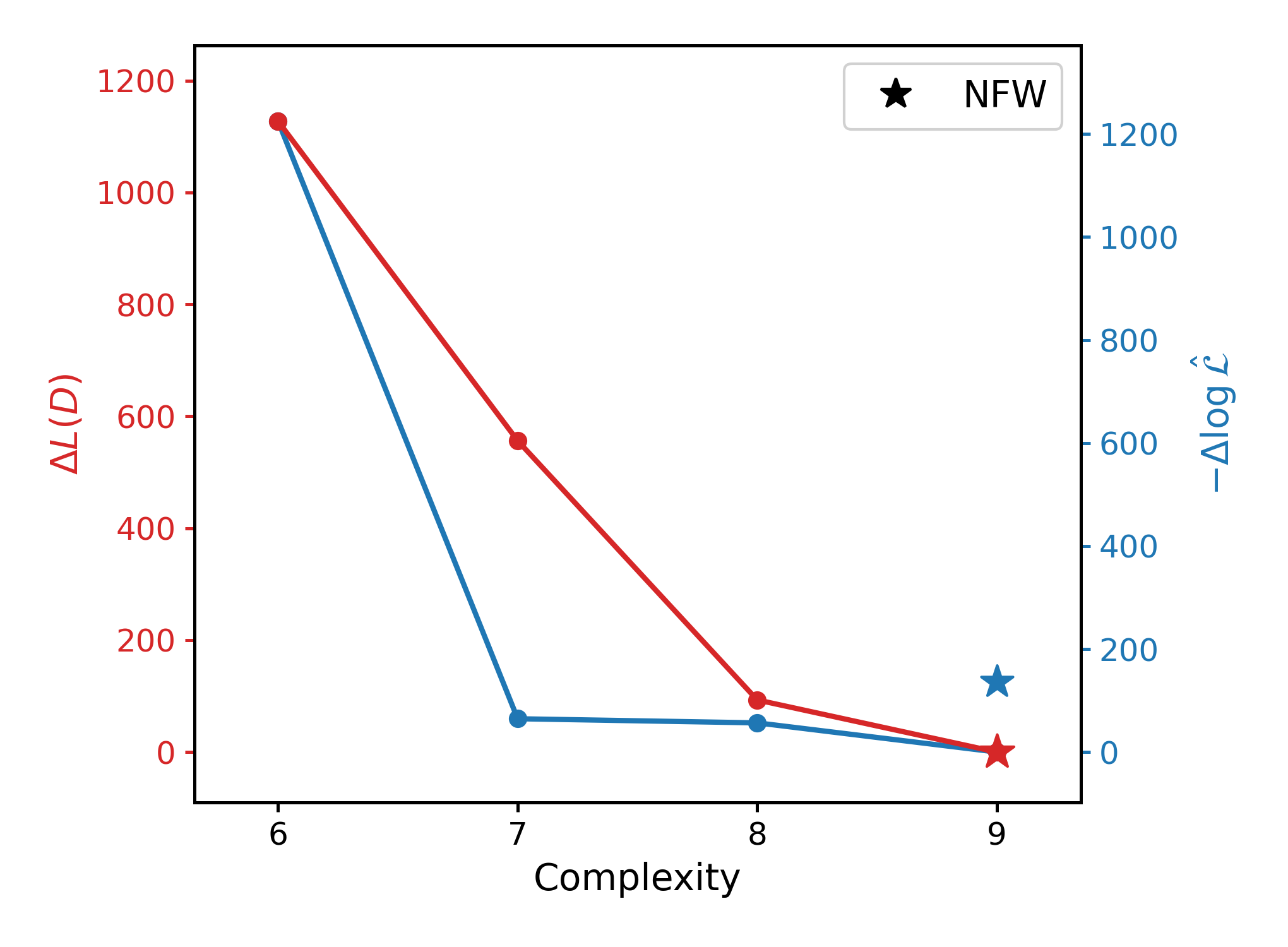}
\caption{
Optimal values of the objectives at each complexity for
a fractional uncertainty of  $f=0.05$. The red curve (left axis) shows the per-complexity minimum description length relative to the global best description-length (NFW in this case). The blue curve (right axis) shows the per-complexity best likelihood relative to the global best likelihood across all models. Stars mark NFW in both metrics. For both axes, lower values on the plot indicate better-performing models.}
\label{fig:pareto_plot}
\end{figure}

The trade-off between model fit (residuals) and complexity is shown in in Fig. \ref{fig:pareto_plot}. The red axis (left, $\Delta L(D)$) shows the total Description Length. The blue axis (right, $|\Delta \log \hat{\mathcal{L}}|$) shows the goodness-of-fit. Both metrics are shown relative to the best-performing model in that metric, so a value of $0$ represents the minimum (best) score. Therefore, a higher point on the plot indicates a worse-performing model.

The NFW profile is marked with a star. It has the best description length, even when it does not have the best likelihood. The function that achieves the best likelihood ($\mathcal{L} = 481.98$ nats) is 


$$\left| \frac{\theta_0}{\theta_2 - r^{\theta_1}} \right|^{\theta_3},$$
but it is penalised for its higher functional complexity, finally ending at rank $545$ in overall $L(D)$.

Fig. \ref{fig:pareto_plot} also shows that the total $L(D)$ is still decreasing at complexity $9$. This suggests that we have not necessarily found the absolute best model, as better-fitting functions may still exist at higher complexities. In fact, all of the top-ranked functions in Table \ref{tab:best_funcs} are already at this complexity $9$ limit. While extending the search to complexities beyond $9$ is a natural next step, it is computationally expensive. We, therefore, present our results as the best-performing functions discovered within the search space up to complexity $9$.

\begin{figure}
\centering
\includegraphics[width=120mm]{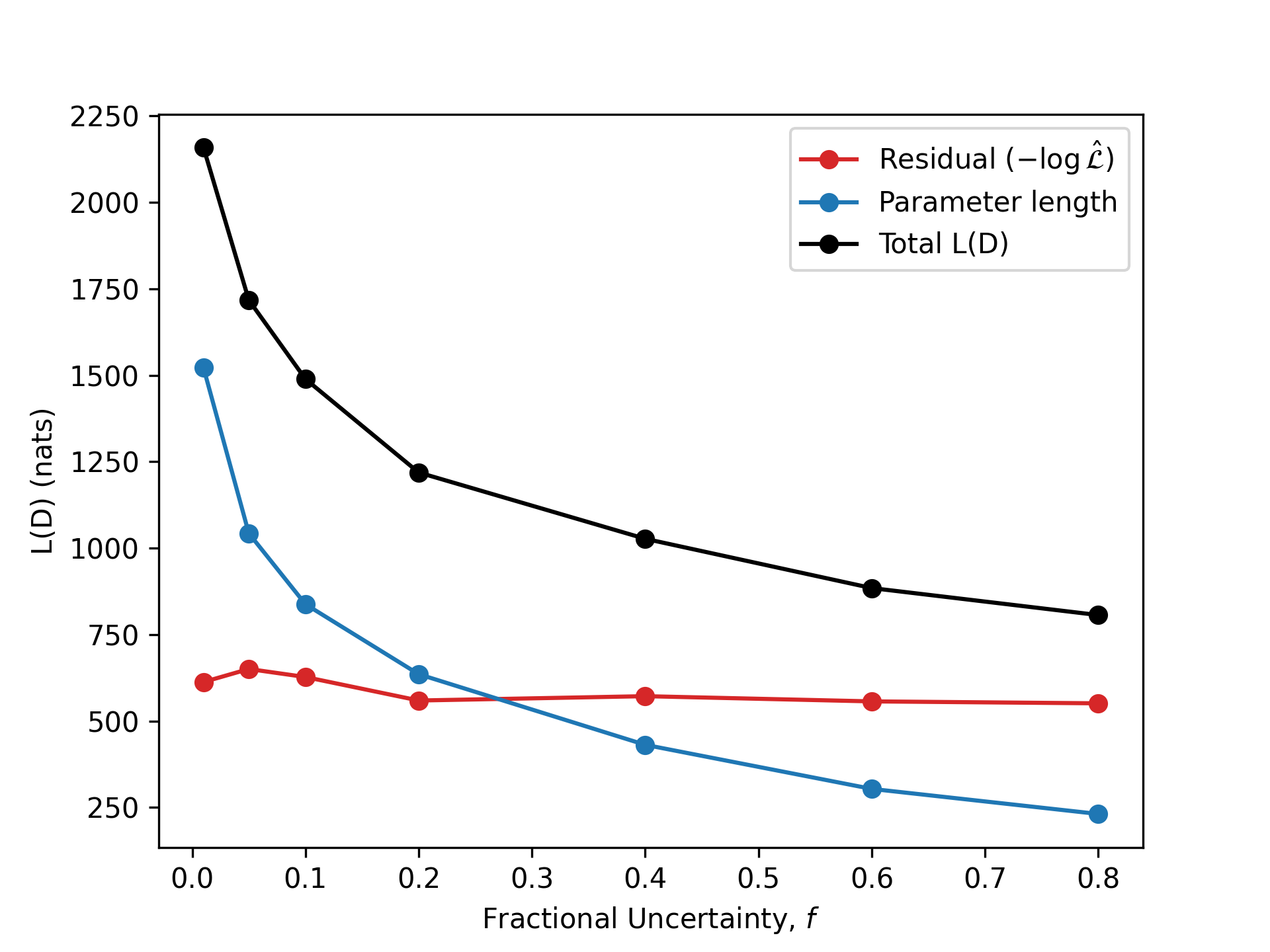}
\caption{Breakdown of the total Description Length ($L(D)$) for NFW as a function of the fractional uncertainty ($f$) in the mock data. 
The Total $L(D)$ (black) is the sum of the 
Residual component ($- \log \hat{\mathcal{L}}$) (red), the 
Parameter length (blue) and the function complexity. The plot shows that the parameter cost (blue line) dominates the total cost at low noise and drops significantly as the data becomes less constraining. The residual term, representing the goodness-of-fit, remains relatively flat at all noise levels.}
\label{fig:nfw_vs_f}
\end{figure}

We repeat this analysis for all fractional uncertainties. Fig. \ref{fig:nfw_vs_f} shows how the different components of the description length change with different fractional uncertainties ($f$) for the NFW profile. The residual component (red line), which represents the goodness-of-fit ($-\log \hat{\mathcal{L}}$), is remarkably flat for all uncertainty levels. This indicates that the model is achieving an optimal fit at each noise level. As the noise $\sigma$ increases, the absolute difference between the data and model also increases, but the normalised residual remains constant. In contrast, the parameter length (blue line) drops significantly as the fractional uncertainty $f$ increases. At low noise, the data is highly constraining, and the parameters are determined with high precision (i.e., narrow posteriors). This high precision represents a large amount of information, which has a high ``cost'' to encode. 

As noise increases, the data's constraining power drops, leading to less precise, low-information (and thus low-cost) parameters. In this high-noise regime, it becomes more common for parameters to be ``snapped'' to zero (i.e., $|\hat{\theta}_i|/ \Delta_i < 1$), which reduces the effective parameter count $p$ for that cluster. For any parameters that are not snapped but are still poorly constrained, the information-dependent part of the parameter cost, $\sum_{a=1}^N \sum_{i=1}^p \log(|\theta_i^{(a)}| / \Delta_i^{(a)})$, tends to zero. Therefore, in the high-noise limit, the parameter length converges to its minimum "structural" cost, which is given by the $p \log 2$ term in Eq. \ref{eq:DL}. For a two-parameter model ($p_{\text{cluster}}=2$) like NFW and our sample of $150$ clusters, this floor is:
\begin{equation*}
    L(\theta)_{\text{min}} = 150 \times 2\log 2 \approx 207.94 \text{ nats},
\end{equation*}
which is presumably the value that the blue line is approaching in Fig. \ref{fig:nfw_vs_f}.

\begin{figure}
\centering
\includegraphics[width=120mm]{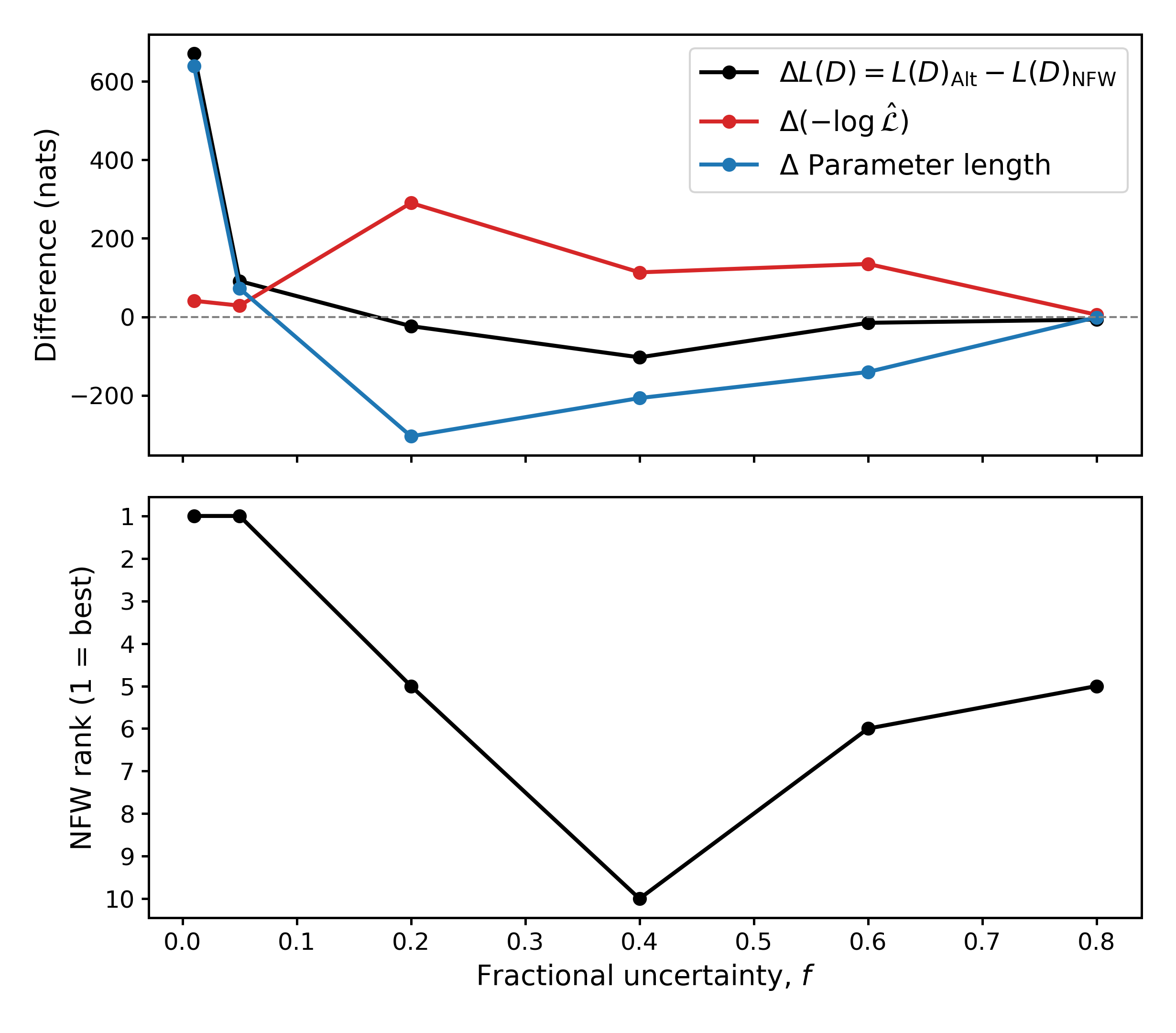}
\caption{The performance of the NFW profile relative to the other ESR-discovered functions, shown as a function of the fractional uncertainty ($f$) in the mock data.\\
\textit{Top panel:} The difference in total Description Length, $\Delta L(D) = L(D)_{\mathrm{Alt}} - L(D)_{\mathrm{NFW}}$, where $L(D)_{\mathrm{Alt}}$ is the description length of the best function other than NFW (black). The equivalent difference for likelihood (red) and parameter length (blue) also shown.
Positive values indicate that NFW is the best-fitting model function. \\
\textit{Bottom panel:} The absolute rank of the NFW profile in the full list of functions, where a rank of $1$ is the best. The plot shows that NFW is the $L(D)$-preferred model at low uncertainty, but it is overtaken by other functions at intermediate $f$. In the high-noise regime ($f \gtrsim 0.6$), it once again 
becomes one of the top-ranked models.}
\label{fig:rank}
\end{figure}

We also studied how the NFW profile's performance compares to the ESR-discovered functions across all data quality levels. Figure~\ref{fig:rank} summarises this analysis. The bottom panel shows NFW's absolute rank (where 1 is best) amongst all ESR functions up to complexity $9$. The top panel shows the $L(D)$ gap, which we define as $\Delta L(D) = L(D)_{\text{alt}} - L(D)_{\text{NFW}}$, where $L(D)_{\text{alt}}$ is the $L(D)$ of the best model found by ESR at each fractional uncertainty (which may differ with $f$). Since a lower $L(D)$ is better, positive values on the black line mean NFW is preferred. The red line shows the difference in fit quality and the blue line shows the difference in parameter cost.

Three regimes appear. At very low noise ($f \le 0.05$), NFW is clearly preferred (rank~1). Both the red and blue curves are positive, showing that NFW achieves a better likelihood and a lower parameter cost than its competitors. As Fig.~\ref{fig:nfw_vs_f} highlights, the parameter-length term dominates the $L(D)$ difference here: some functions can approach NFW’s likelihood, but only at the price of substantially finer parameter tuning.

However, in the intermediate-noise regime ($f \approx 0.2 - 0.4$), NFW's performance gets worse. Here, the red line is still positive (NFW fits better), but the blue line is strongly negative. This indicates that an alternative function has a lower parameter cost. Actually, the preferred model for $f = {0.2, 0.4, 0.6}$ is $\theta_0 / (r^2(r+1))$. The MDL framework concludes that NFW's slightly better fit is not worth its much higher parameter "fine-tuning" cost, so it prefers the alternative.

At very large $f$, both likelihood differences and parameter costs compress, so NFW may no longer be best but typically remains competitive. This behaviour is likely driven by the fractional-uncertainty model used in our mock. We set $\sigma(r) = f \times \rho(r)$, which mirrors weak lensing–like errors that scale with the signal amplitude. This choice makes the absolute uncertainties at small radii bigger than those in the outskirts, so the outer data points carry more weight in the likelihood. At large $r$, NFW asymptotes to $r^{-3}$. Many alternative profiles, like $\theta_0 / (r^2(r+1))$, can reproduce this outer slope while differing in the centre. As a result, the fractional model effectively down-weights the inner cusp ($r^{-1}$) that distinguishes NFW, making alternative functions competitive in likelihood. Therefore, more simple functions that can reproduce the outer behaviour but have less parameters are preferred, which is exactly what we are seeing in Fig. \ref{fig:rank}.

This is just a property of WL-like weighting. In practice, one could combine WL with probes that better constrain the inner regions (e.g. strong lensing or stellar kinematics). This also means model recoverability depends as much on the as on the fitting criterion.

It is interesting to note, however, that even when NFW is not the $L(D)$-winner, it always remains one of the top models. As the bottom panel of Figure~\ref{fig:rank} shows, NFW remains within the top 10 models across all uncertainty levels. This suggests that if NFW is the best description of the data, ESR should be able to recover it as one of the best-fitting models.

    In addition to noise, another important factor that influences model selection is the amount of available data. To study this effect, we assess how the description length changes under variations in the number of clusters $N$. After running the analysis for $N=150$ clusters, we extrapolate our results to arbitrary sample sizes. The only terms that depend on $N$ in Eq.~\ref{eq:DL_global} are the total likelihood $\sum^N \log \mathcal{L}$ and the parameter information term $\sum^N\sum^p (\frac{1}{2} \log(I_{ii}) + \log(|\theta_i|)$. Both of these terms scale linearly with $N$, allowing for a trivial transformation. 
    
We can now use this framework to quantify how the statistical preference for the NFW profile scales with data quantity (i.e., the number of clusters, $N$). Since the extrapolation to a different number of clusters is linear, we can only study regimes for which NFW is identified already as the best model. So we focus on the two low-noise regimes, $f = 0.01$ and $f = 0.05$, which are the cases where Fig.\ref{fig:rank} showed NFW is the best-performing model for the full sample.

\begin{figure}
\centering
\includegraphics[width=120mm]{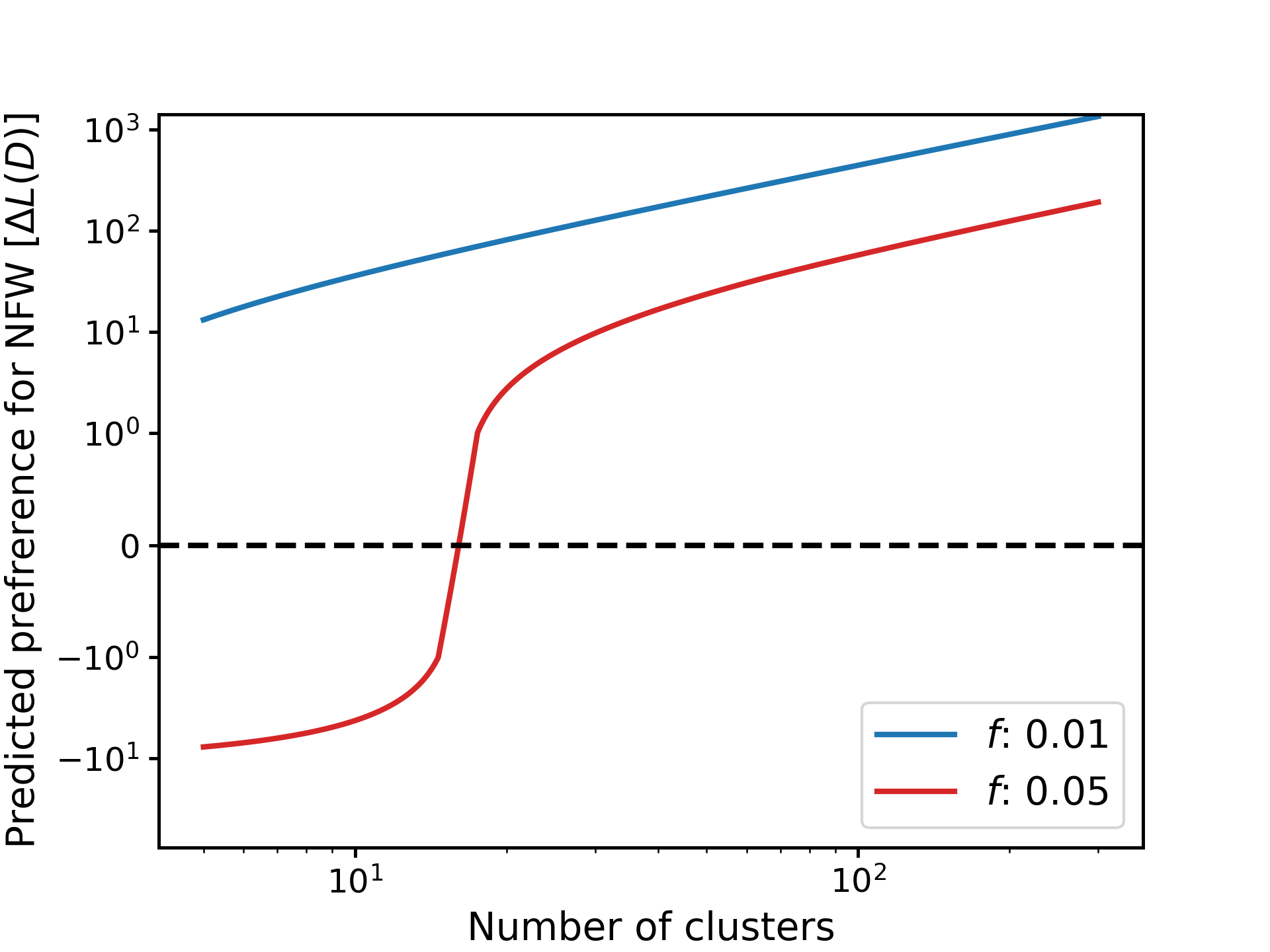}
\caption{Change in description length ($\Delta L(D)$) between NFW and the best alternative dark matter halo model. The curves show results for $2$ different levels of Gaussian noise in the ESD data: $f \in \{0.01, \, 0.05\}$ as a function of the number of clusters. Any point higher than $0$ shows a preference for NFW.}
\label{fig:preference}
\end{figure}

For these two noise levels, we compute the description length difference, defined as $\Delta L(D) = L(D)_{\text{alternative}} - L(D)_{\text{NFW}}$, where $L(D)_{\text{alternative}}$ is the $L(D)$ of the next-best function. This difference is plotted as a function of $N$ in Fig.\ref{fig:preference}. With this definition, positive values of $\Delta L(D)$ indicate a preference for NFW.

The results show that for the very low-noise case ($f = 0.01$), the preference for NFW is strong even with a very small number of clusters. As the uncertainty increases to $f = 0.05$, a larger dataset is required. While NFW is still preferred, a sample of at least $N \approx 20$ clusters is needed to accumulate significant statistical evidence. 

At the next noise level we studied, $f=0.20$, NFW was no longer the top-ranked model (falling to rank $5$) for our $N=150$ sample. It is possible that NFW could still be recovered with $N=150$ clusters at an intermediate noise level (e.g., $f \approx 0.10$), or that recovering NFW at $f=0.20$ would just require a larger sample size ($N > 150$). We leave a detailed exploration of this for future work.

\section{Conclusions and Future Work}
\label{sec:conclusion}

\subsection{Overview}

We have demonstrated how Exhaustive Symbolic Regression (ESR) can be applied to constrain dark matter density profiles, specifically focusing on recovering halo structures from weak lensing Excess Surface Density (ESD) measurements.

To illustrate the method, we applied ESR to a mock sample of galaxy clusters with ESD profiles generated from the standard NFW model. We modelled the observational uncertainty as a constant fractional error, $\sigma_i = f\,\Delta\Sigma_{\rm true}(R_i)$, and tested the pipeline across six distinct noise levels: $f \in \{ 0.01, \, 0.05, \, 0.2, \, 0.4, \, 0.6, \, 0.8 \}$. By running ESR on this set of mocks, we studied the ranking of best-fitting functions to determine the conditions under which the input NFW profile is successfully recovered.

In the low-noise cases ($f= 0.01$ and $f= 0.05$), ESR identifies NFW as the optimal model. In this regime, the data is precise enough to penalise deviations from the true profile shape, and very few alternative profiles can reproduce the signal without requiring finely tuned parameters.

In the intermediate and high-noise regimes ($f \geq 0.2$), NFW is typically not recovered as the rank-$1$ model, though it consistently remains within the top $10$. This result is probably a consequence of the fractional uncertainty modelling. Because the error scales with the signal ($\sigma_i \propto \Delta\Sigma(R_i)$), the absolute uncertainty is smallest at large radii. Consequently, the likelihood calculation is dominated by the cluster outskirts, where the profile asymptotes to $r^{-3}$. Simpler functions that can mimic this outer slope can achieve a likelihood comparable to NFW but with a lower parameter cost, thus outranking NFW in $L(D)$.

Nevertheless, the fact that NFW remains a leading candidate (top 10) even at higher noise levels is encouraging. It indicates that for current surveys with precisions similar to HSC ($f \approx 0.6$), ESR should successfully recover NFW as one of the top models if the underlying physics is indeed NFW-like.

We also investigated the impact of sample size by extrapolating our results from $N=150$ clusters. For the low-noise cases ($f \le 0.05$), we found that the preference for NFW is statistically preferred even with a very small number of clusters. At higher noise levels, larger samples would be required to overcome the preference for simpler functions driven by the outer-slope weighting.

Future work could extend this analysis in several directions. One could probe higher model complexities (beyond complexity 9) to assess the stability of these rankings. Additionally, more realistic physical modelling could be incorporated.

For instance, by not explicitly modelling the ICM gas, our inference effectively constrains the total mass profile rather than the purely dark-matter contribution. A more complete treatment would require marginalizing over the gas distribution (using observed or empirical ICM profiles) to isolate the dark matter profile, or using mock data to quantify any bias introduced by its neglect.

Similarly, we have assumed the dark matter halos are spherically symmetric, but in dark matter simulations halos are found to be strongly triaxial, with moderate triaxiality is found in observations \cite{Chiu_2018}. Studies have found  for a sufficiently large sample of randomly oriented samples the bias in parameters when fitting an NFW halo is small. Hence it seems to first order it should be possible to model this triaxiality as an additional noise term. This could be studied in detail in mocks using a suite of functions with triaxial scale parameters. With future developments to ESR algorithm, and larger datasets, in the future it may be possible to constrain three dimensional halo parameterisations. 

As data quality improves in future surveys down to the 5\% level we found necessary to recover NFW with $< 150$ clusters, properly treating systematic effects will certainly be vital to infer the correct halo profile in data, and a detailed study of potential model misspecification in mocks in this regime should be conducted in the future.

More broadly, the framework we have developed is applicable to any dataset that constrains the radial distribution of dark matter in galaxies or galaxy clusters. The core idea is straightforward: use symbolic regression to generate candidate expressions for the density profile $\rho(r)$, compute the corresponding observables, and rank the models using the MDL principle. This methodology could be applied, for instance, to galactic rotation curves.

\subsection{Application to rotation curves}

The gold standard for recovering the dark matter distribution around individual galaxies is rotation curves, the rotational velocity of an observable tracer (stars or gas) as a function of radius. Spiral galaxies are particularly good systems to study this as they are approximately axisymmetric and their stars and gas clouds follow nearly circular orbits hence directly tracing the gravitational potential. Due to the linearity of Poisson's equation, the total circular velocity can be expressed as the quadrature sum of the circular velocity contributed by each mass component

\begin{eqnarray}
v^2_{c}(r) = v^2_{\text{baryonic}}(r) + v^2_{\text{DM}}(r),
\label{eq:rot_vel}
\end{eqnarray}
where $v_{\text{DM}}(r)$ is the contribution of dark matter, and $v_{\text{baryonic}}(r)$ the baryonic component, given by

\begin{eqnarray}
v^2_{\text{baryonic}}(r) = \Upsilon_{\text{disc}}v^2_{\text{disc}}(r) + \Upsilon_{\text{bulge}}v^2_{\text{bulge}}(r) + \Upsilon_{\text{gas}}v^2_{\text{gas}}(r),
\end{eqnarray}
which includes contributions from the disk, bulge and gas, each weighted by their respective mass-to-light ratios $\Upsilon_{x}$.

We can use the functions generated by ESR as candidate dark matter density profiles $\rho_{DM} (r)$. These can be integrated to give a mass and a velocity profile
\begin{align}
M_{\text{DM}}(r) &= 4 \pi \int_0^r {r^\prime}^2 \rho_{\text{DM}}(r^\prime)\, dr^\prime, &
v^2_{\text{DM}}(r) &= \frac{G M_{\text{DM}}(r)}{r}
\end{align}

To fit to observational data, one needs to combine the dark matter model with the baryonic contribution to match the total circular velocity (Eq.~\ref{eq:rot_vel}). 
One can infer the baryonic matter distribution (and thus its contribution to the circular velocity) from measurements of the stellar and gas velocities, typically based on photometry and HI observations\cite{lelli2016sparc}.
The remaining, unaccounted for contribution to the observed $v_{\rm c}$ is due to the dark matter, allowing one to infer its density profile.
These measurements in turn depend on galaxy-specific parameters such as mass-to-light ratios $\Upsilon$, the galaxy distance $D$, and the disk inclination $i$. This introduces additional complexity, as fitting the combined model requires simultaneous optimisation of both the function parameters describing the dark matter profile and galaxy-specific quantities associated with the baryonic component. As a result, the optimisation becomes a significantly more challenging task, involving exploration of a higher-dimensional parameter space.




Still, rotation curves provide a complementary probe to lensing, especially at smaller galactic radii where lensing is less sensitive. It would be interesting to test how closely dark matter density profiles obtained from weak lensing match those obtained from rotation curves. It should also be straightforward to analyse both datasets combined. ESR naturally accounts for differences in the number of data points and measurement precision, one simply needs to combine description lengths carefully: model complexity should only be penalised once, while the likelihood contributions from each dataset are summed, in a similar way to what was done here for multiple galaxy clusters. Together, these two methods can provide a data-driven approach to constraining dark matter profiles. This might help reveal the underlying physical properties of dark matter and guide the development of alternative dark matter models as the quantity and quality of available data continue to grow.

\section*{Data availability}

The data underlying this article will be shared on reasonable request to the corresponding author.

\ack{
We thank the referees for their careful reading of the manuscript and for their valuable suggestions. TY acknowledges the support of a UKRI Frontiers Research Grant [EP/X026639/1], which was selected by the
European Research Council. 
DJB was supported by the Simons Collaboration on ``Learning the Universe’’ and acknowledges that support was provided by Schmidt Sciences, LLC.
HD is supported by a Royal Society University Research Fellowship (grant no. 211046).
PGF acknowledges support from STFC and the Beecroft Trust. 
This work was performed using resources provided by the Cambridge Service for Data Driven Discovery (CSD3) operated by the University of Cambridge Research Computing Service (www.csd3.cam.ac.uk), provided by Dell EMC and Intel using Tier-2 funding from the Engineering and Physical Sciences Research Council (capital grant EP/P020259/1), and DiRAC funding from the Science and Technology Facilities Council (www.dirac.ac.uk).}


\bibliographystyle{RS} 
\bibliography{refs}  

\end{document}